\documentclass[conference]{IEEEtran}

\ifCLASSINFOpdf
\else
   \usepackage[dvips]{graphicx}
\fi
\usepackage{url}

\usepackage{graphicx}

\usepackage{algorithm, algorithmic} 
\usepackage{float}  
\usepackage{lipsum}

\usepackage{diagbox}
\usepackage{xcolor}

\usepackage{amsmath,amssymb,amsfonts,mathrsfs,bm}  
\usepackage{graphicx}   
\usepackage{amsthm} 
\usepackage{setspace}
\usepackage{subfigure}
\usepackage{enumerate}

\usepackage{verbatim}  
\usepackage{array} 

\usepackage{multirow}

\usepackage{hyperref}
\hypersetup{
colorlinks=true,
linkcolor=black,
anchorcolor=black,
filecolor=black,
urlcolor=black,
citecolor=black,}

\usepackage[capitalize]{cleveref} 

\newtheorem{theorem}{Theorem}

\newtheorem{lemma}{Lemma}

\newtheorem{assumption}{Assumption}

\floatname{algorithm}{Procedure}

\newcommand{\Real}{\mathbb{R}}

\newcommand{\Complex}{\mathbb{C}}

\newcommand{\calE}{\mathcal{E}}

\newcommand{\calG}{\mathcal{G}}

\newcommand{\calI}{\mathcal{I}}
\newcommand{\calJ}{\mathcal{J}}

\newcommand{\calS}{\mathcal{S}}
\newcommand{\calT}{\mathcal{T}}

\newcommand{\calV}{\mathcal{V}}

\newcommand{\bD}{\mathbf{D}}
\newcommand{\be}{\mathbf{e}}
\newcommand{\bE}{\mathbf{E}}

\newcommand{\bF}{\mathbf{F}}

\newcommand{\bL}{\mathbf{L}}

\newcommand{\bQ}{\mathbf{Q}}

\newcommand{\bU}{\mathbf{U}}

\newcommand{\bV}{\mathbf{V}}

\newcommand{\bW}{\mathbf{W}}

\newcommand{\bX}{\mathbf{X}}

\newcommand{\scP}{\mathscr{P}}

\newcommand{\bSigma	}{\bm{\Sigma}}

\usepackage{booktabs} 
\usepackage{makecell}

\usepackage{geometry}
\begin{document}

\title{Subset Random Sampling of Finite Time-vertex Graph Signals}

\author{\IEEEauthorblockN{Hang Sheng$^\ast$, Qinji Shu$^\ast$, Hui Feng$^{\ast,\dagger}$, Bo Hu$^{\ast,\dagger}$}

\IEEEauthorblockA{$^\ast$ School of Information Science and Technology, Fudan University, Shanghai 200433, China\\
$^\dagger$ State Key Laboratory of Integrated Chips and Systems, Fudan University, Shanghai 200433, China\\
hsheng20@fudan.edu.cn, qjshu22@m.fudan.edu.cn, \{hfeng, bohu\}@fudan.edu.cn
}}

\maketitle

\begin{abstract}
Time-varying data with irregular structures can be described by finite time-vertex graph signals (FTVGS), which represent potential temporal and spatial relationships among multiple sources. While sampling and corresponding reconstruction of FTVGS with known spectral support are well investigated, methods for the case of unknown spectral support remain underdeveloped. Existing random sampling schemes may acquire samples from any vertex at any time, which is uncommon in practical applications where sampling typically involves only a subset of vertices and time instants. In sight of this requirement, this paper proposes a subset random sampling scheme for FTVGS. We first randomly select some rows and columns of the FTVGS to form a submatrix, and then randomly sample within the submatrix. Theoretically, we prove sufficient conditions to ensure that the original FTVGS is reconstructed with high probability. Also, we validate the feasibility of reconstructing the original FTVGS by experiments.

\end{abstract}

\IEEEpeerreviewmaketitle

\section{Introduction}

Graphs can capture the underlying structure of data, making graph signals valuable for processing spatial information of data \cite{2023graph}. When data exhibit temporal variations, signal analysis also involves discussions in the time dimension. The concept of time-vertex graph signals (TVGS) was introduced in \cite{timevertex}. For example, Grassi \emph{et al.} formed a product graph by stacking graph signals in finite time instances \cite{timevertex}, which we refer to as finite time-vertex graph signals (FTVGS).

Sampling and reconstruction for TVGS is a crucial aspect, aiming to recover the entire signal from partial observations, thereby saving on the costs of acquisition, storage, and processing \cite{sheng2024sampling}. TVGS can be reconstructed from samples because real-world TVGS typically exhibit low-frequency characteristics in both the vertex and time domains. For example, when representing FTVGS with a matrix, this matrix would be low-rank and smooth.

There are some advancements in the research of sampling of FTVGS, provided that the joint time-vertex Fourier transform (JFT) spectrum support of FTVGS is known. The separate sampling scheme applies sampling to the graph and time domains of FTVGS respectively \cite{sampling2018}, while the joint sampling scheme further reduces the sampling rate based on the JFT spectrum of FTVGS \cite{sheng2024sampling}. The sampling framework based on subspace prior information applies the sampling framework of shift-invariant spaces to FTVGS \cite{2024sample}. Signal sampling and reconstruction algorithms based only on known graph spectral support were also proposed \cite{2023jointsamp}.

However, in reality, it is challenging to know the JFT spectrum of the FTVGS to be sampled beforehand. For instance, in FTVGS acquired by sensor networks, each row of the matrix represents a time series collected by a sensor, and each column represents a traditional graph signal. Before deploying the sensors, it is impossible to know the specific information of the FTVGS before sampling. Thus, the sampling methods for FTVGS based on known spectrum are not feasible.

To address the challenges posed by the unknown spectrum, we adopt a random sampling strategy. One approach is to sample the entire FTVGS through some unbiased stochastic process, which assumes that all elements in the matrix have an equal probability of being sampled \cite{MC1,MC2}. However, this assumption implies all the vertices could be sampled at any time (\cref{fig:area} (a)), which does not meet the requirement of reducing the number of sensors to be sampled.
Cai \emph{et al.} proposed a more flexible sampling scheme \cite{CCS}, but it still requires every row and column of the matrix to produce samples (\cref{fig:area} (b)). In practice, due to hardware, environmental, or cost limitations, we prefer to sample only a subset of rows and columns.

We propose a subset random sampling scheme for FTVGS to meet practical needs, significantly reducing the number of rows and columns from which samples are produced (\cref{fig:sub_samp} (c)). Specifically, we first randomly select subsets of rows and columns of the matrix to form a submatrix, and then randomly observe samples within the submatrix. Theoretically, under common assumptions, we prove sufficient conditions to ensure a high probability of reconstructing the original FTVGS. Additionally, the subset random sampling scheme impacts the complexity of the required samples and the accuracy of reconstructing the original signal, which we discuss in detail in \cref{sec:sample,sec:experiment}.

\section{Related Works}

The representation of FTVGS is a matrix, so the random sampling for FTVGS is also a random sampling problem on matrices. The pioneering work \cite{MC1} studied the matrix completion (MC) problem based on the Bernoulli sampling model and proved a lower bound on the number of samples sufficient for exact recovery with high probability. Subsequent research on MC has extended beyond Bernoulli sampling to include uniform sampling with and without replacement \cite{MC2,noisyMC}. 

In subsequent studies, uniform sampling in MC was combined with CUR decomposition, resulting in a more flexible cross-concentrated sampling (CCS) \cite{CCS}. But, in CCS, at least one element from each row or column of the FTVGS must still be sampled.

\begin{figure}[htbp]
    \begin{center}
        \includegraphics[scale=0.45]{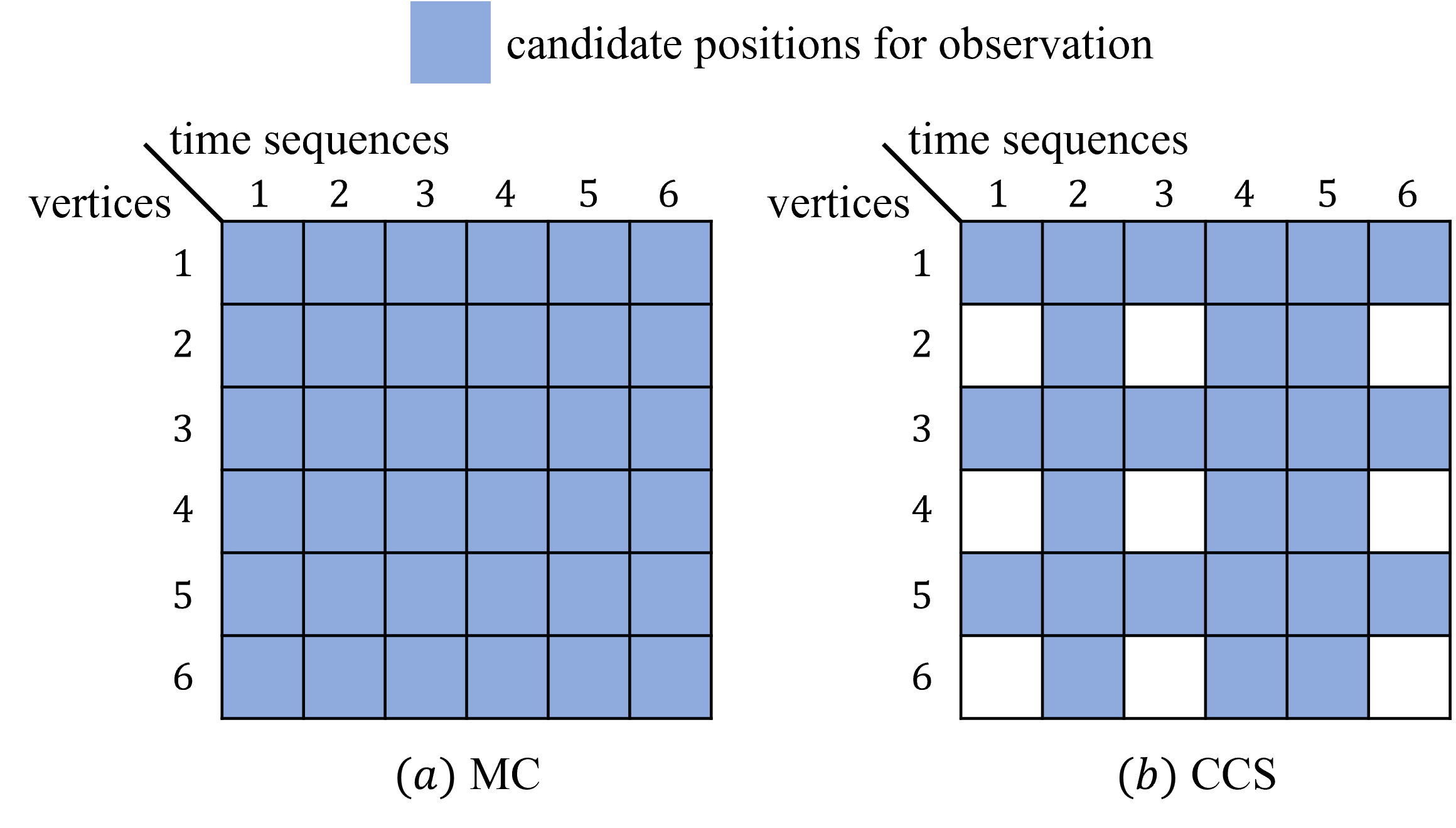}   
    \end{center}
    \caption{The visualization of candidate positions for observation.}
    \label{fig:area}
\end{figure}

\cref{fig:area} visualizes the potential sampling regions for uniform sampling in MC and CCS. It is evident that neither method reduces the number of rows and columns being sampled in the matrix.

We aim to restrict the sample indices within subsets of the rows and columns of FTVGS, ensuring no samples are produced on unselected rows and columns. This approach is motivated by practical applications, where sampling is often desired on only partial vertices and time instants. For example, in sensor networks, we want to deploy as few sensors as possible, and it is not supported by hardware for each instant to have an equal probability of being sampled. In social networks, not everyone is willing to share their thoughts, and there may be times when participants cannot be recruited.

\section{Model}
\label{sec:model}

An undirected graph can be represented as $\calG = (\calV_\calG, \calE_\calG)$, where $\calV_\calG$ is the set of vertices with $|\calV_\calG| = N$, $\calE_\calG$ is the set of edges. We consider every vertex in $\calV_\calG$ relates to a time sequence of length $T$, which can be represented by a directed cyclic graph $\calT = (\calV_\calT, \calE_\calT)$. Such signals are FTVGS \cite{timevertex}, whose graphical model is $ \calG \times \calT$, where $\times$ is the Cartesian product. An FTVGS is $\bX \in \Complex^{N \times T}$. Each row of $\bX$ is a finite time sequence on a vertex, and each column is a static graph signal. 

Suppose each edge of $\calG$ is assigned an orientation, which is arbitrary but fixed. We then have the incidence matrix $\bQ \in \Real^{N \times M}$ and the graph Laplacian matrix $\bL_\calG = \bQ \bQ^T$. Then the gradient on $\calG$ can be defined as $\bQ^T \bX$ and the graph total variation at time $j$ is $\bX(:, j)^T \bL \bX(:, j)$.

Then we denote first and second-order temporal differential operators by $\bD_1 \in \Real^{T \times (T-1)}$ and $\bD_2 \in \Real^{T \times (T-2)}$. For each $i \in \calV_\calG$, the first and second-order differences of the signal on $\calT$ can be represented as $\bX \bD_1(i,j)$ and $\bX \bD_2(i,j)$. The gradient of $\bX$ in the graph and time domains is
\begin{equation*}
    \triangledown \bX(i,j) = \left[\begin{matrix}
    \bQ^T \bX(i,j) \\
    \bX \bD_1(i,j)
    \end{matrix} \right].
\end{equation*}
These are widely used assumptions.
\begin{assumption}
\label{as_r}
    An FTVGS $\bX \in \Complex^{N \times T}$ is a rank-$r$ matrix, where $r< \min \{ N,T \}$. For all $i \in \calV_\calG, j \in \calV_\calT$, 
    $$\max \left\{ \left| \bX(:,j)^T \bL \bX(:,j) \right|, \parallel \triangledown \bX(i,j) \parallel, \left| \bX \bD_2(i,j) \right| \right\} \leq C $$
    for some constant $C$.
\end{assumption}

The low rank implies that $\bX$ can be expressed as a matrix with only $r$ non-zero rows (or columns) through a row (or column) linear transformation. This is consistent with the assumption of bandlimited JFT spectra of FTVGS, as JFT is a type of linear transformation. That is, $\bX = \Psi_\calG \bF_\calG, \bX^T = \Psi_\calT \bF_\calT$, where $\Psi_\calG$ and $\Psi_\calT$ are the graph Fourier transform (GFT) basis matrix and discrete Fourier transform (DFT) basis matrix, respectively \cite{timevertex}. $\bF_\calG$ and $\bF_\calT$ are the GFT and DFT spectra respectively, where each $\bF_\calG(:,j)$ is the bandlimited spectrum of the graph signal $\bX(:,j)$ and each $\bF_\calT(:,i)$ is the bandlimited spectrum of the time sequence $\bX(i,:)$ \cite{sheng2024sampling}.

The graph total variation on $\calG$, second-order difference on $\calT$, and gradient are all bounded implies that $\bX$ is smooth. That is, there are no abrupt changes in $\bX$ in either spatial or temporal domains. Correspondingly, in the context of TVGS processing theory, the smoothness of $\bX$ somehow indicates that its JFT spectrum exhibits low-frequency characteristics.

Since $\bX$ is rank-$r$, its thin singular value decomposition (SVD) is
$$ \bX = \bU \bSigma \bV^T, $$
where $\bU \in \Real^{N \times r}$, $\bSigma \in \Real^{r \times r}$ and $\bV \in \Real^{T \times r}$. 

\begin{assumption}
\label{as_mu}
    \cite{MC1,MC2} An FTVGS $\bX \in \Complex^{N \times T}$ satisfying \cref{as_r} is $\{ \mu_1, \mu_2 \}$-incoherence for some constants $1 \le \mu_1(\bX) \le \frac{N}{r}, 1 \le \mu_2(\bX) \le \frac{T}{r}$ provided
    $$ || \bU ||_{2,\infty} \le \sqrt{\frac{\mu_1(\bX) r}{N}}, || \bV ||_{2,\infty} \le \sqrt{\frac{\mu_2(\bX) r}{T}}, $$
    where $|| \bU ||_{2,\infty} := \max_i \left(\sum_j (\bU(i,j))^2 \right)^{1/2}$.
\end{assumption}

Coherence indicates whether the energy of $\bX$ is uniformly distributed across columns and rows. Low coherence means that the structure of $\bX$ is relatively uniform, making it easier to obtain uniform samples during random sampling.

\section{Method}
\label{sec:sample}

\begin{figure}[htbp]
    \centering
    \subfigure[Randomly select rows and columns]
    {
	    \includegraphics[scale=0.45]{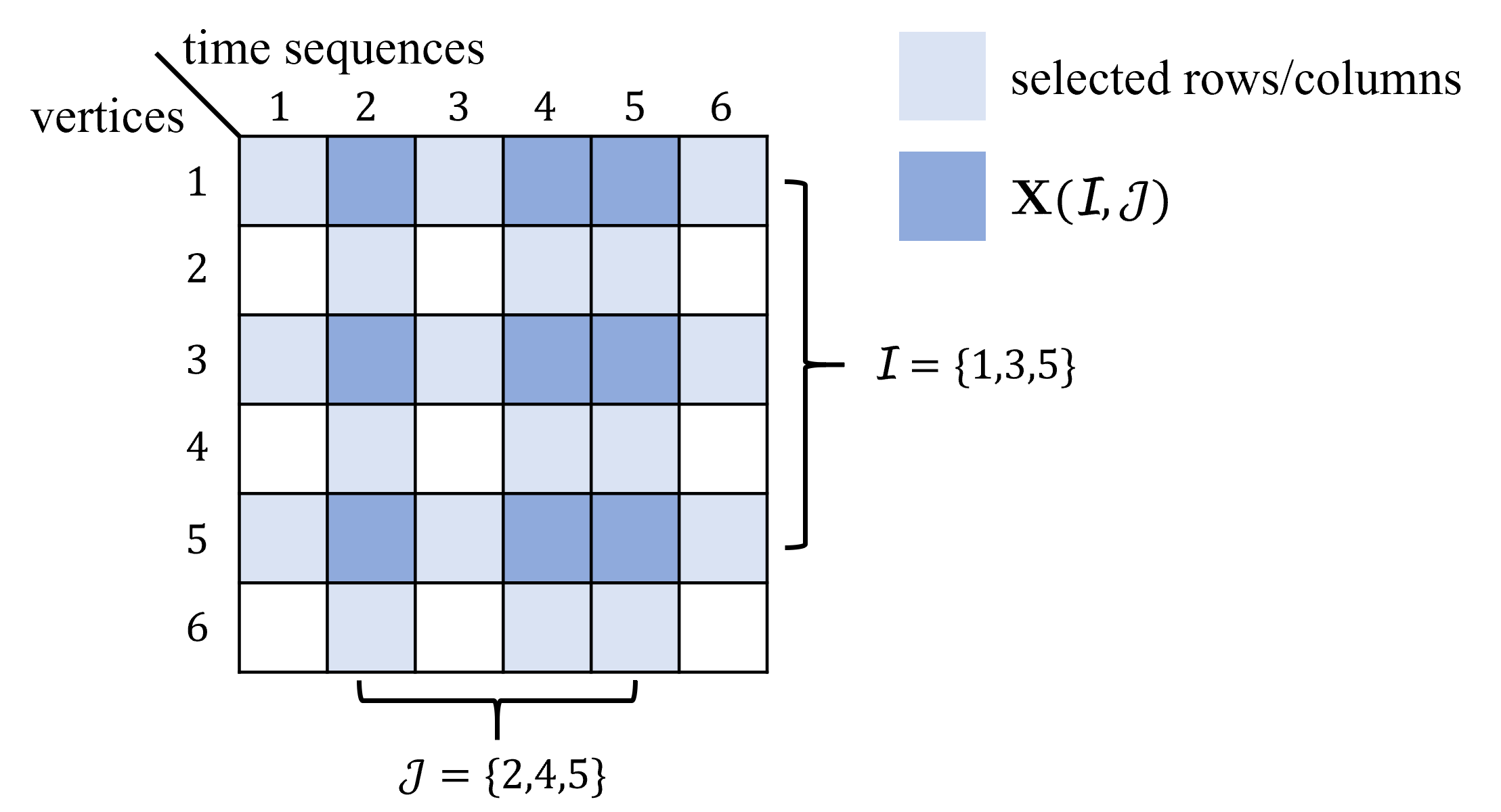}   
	} \\
 
    \subfigure[Sampling on $\bX(\calI, \calJ)$]
    {
	    \includegraphics[scale=0.45]{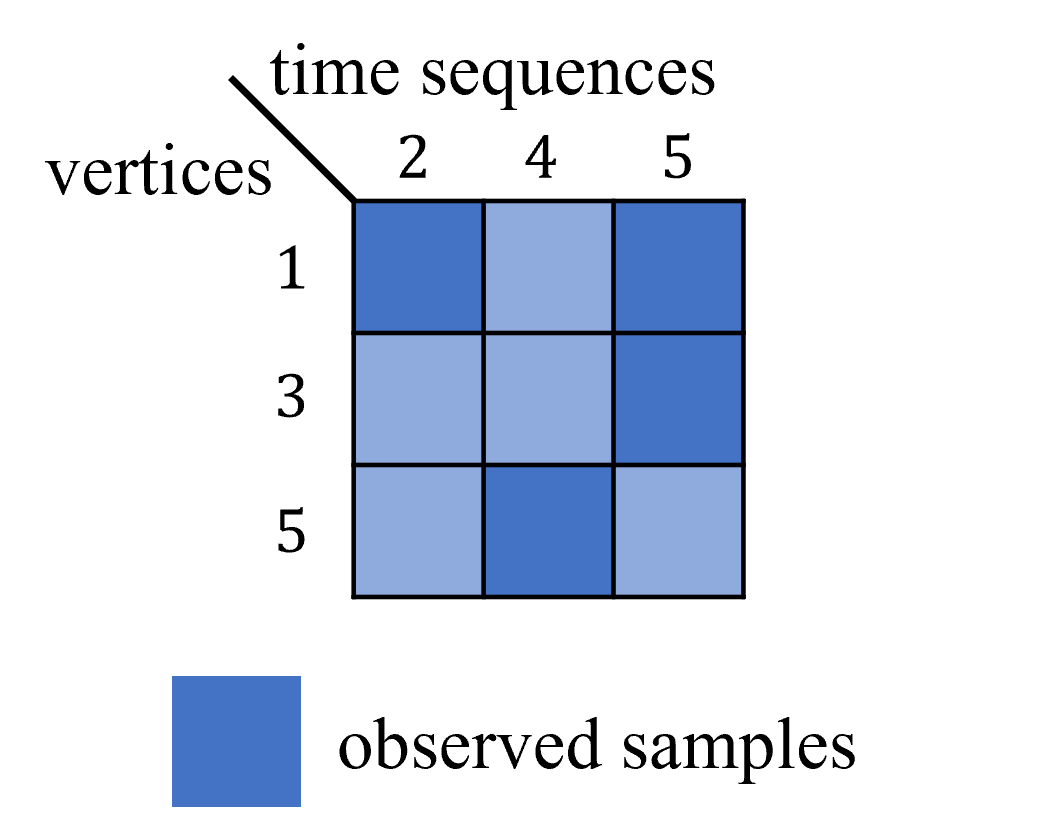}   
    }
    \subfigure[Position of the samples in $\bX$]
    {
	    \includegraphics[scale=0.45]{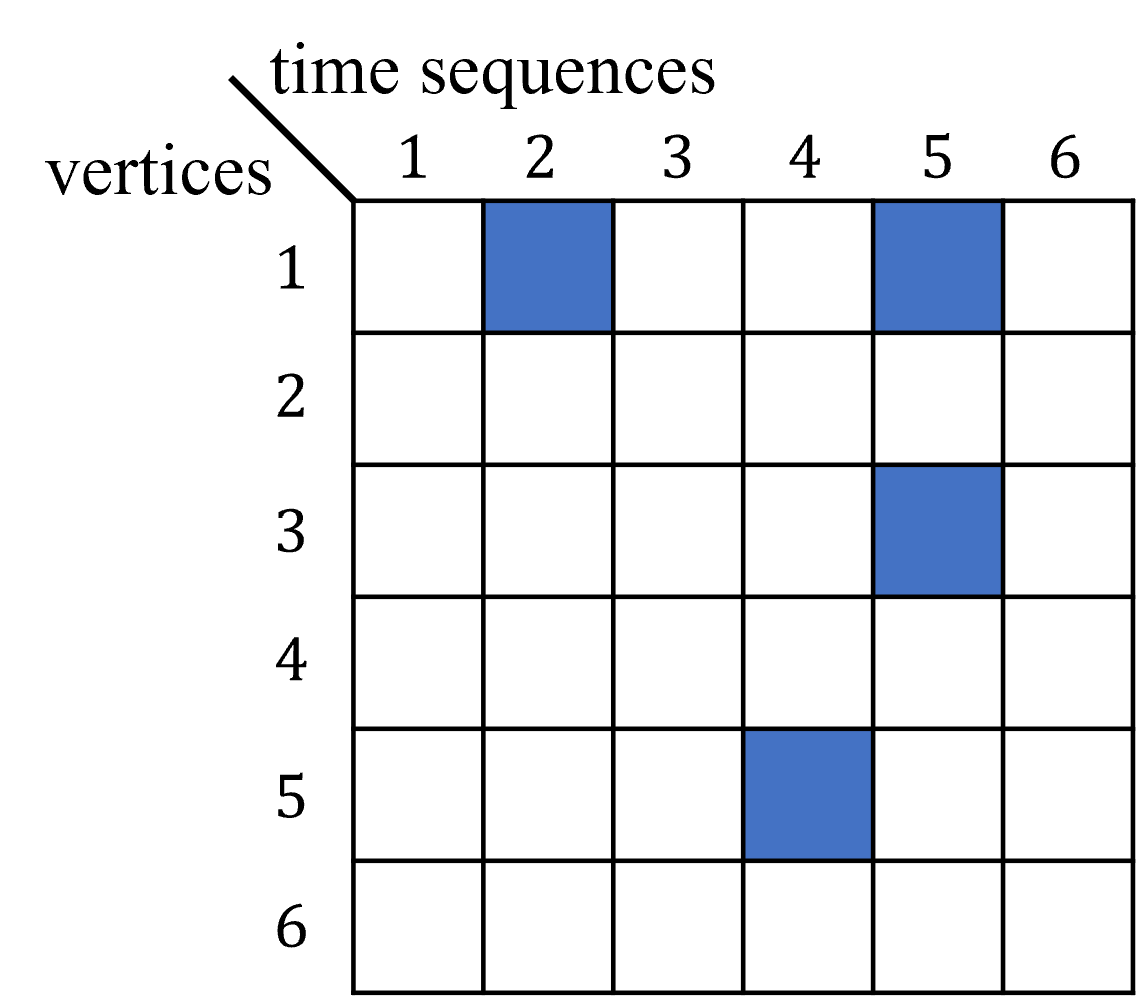}   
    }
    \caption{Flowchart of the subset random sampling scheme. }
\label{fig:sub_samp} 
\end{figure}

We propose \textit{a subset random sampling scheme} for signals, as illustrated in \cref{fig:sub_samp}. Formally, we first randomly select a subset of rows and columns from the FTVGS without replacement (\cref{fig:sub_samp} (a)), \textit{i.e.}, 
$$\calI \subseteq \calV_\calG, \calJ \subseteq \calV_\calT.$$
Let $\bX(\calI, \calJ)$ be the submatrix of $\bX$ composed of rows indexed by $\calI$ and columns indexed by $\calJ$. Let the samples observed from FTVGS $\bX$ be indexed by a set $\calS$, and $|\calS|$ denote the number of samples. Subsequently, we uniformly observe samples within the submatrix $\bX(\calI, \calJ)$ with replacement (\cref{fig:sub_samp} (b), (c)), \textit{i.e.},
$$ \calS = \{ (i, j): i \in \calI, j \in \calJ \}. $$
Let $\scP_\calS(\cdot)$ be the projection operator onto $\calS$.

Unlike existing methods, we would like to sample randomly on the submatrix $\bX(\calI, \calJ)$ instead of $\bX$, ensuring that no samples are observed on unselected rows and columns. We summarize the sampling process as Procedure \ref{proc:samp}.

\begin{algorithm}[htbp] 
\caption{Subset random sampling}
\label{proc:samp}
\begin{algorithmic}[1]

\REQUIRE $\bX$

\STATE Uniformly sample row indices $\calI$ without replacement, yielding $\bX_R = \bX(\calI, :)$.

\STATE Uniformly sample column indices $\calJ$ without replacement, getting $\bX_{RC} = \bX(\calI, \calJ)$.

\STATE Uniformly sample entries in $\bX_{RC}$ and get $\calS$.

\ENSURE $\scP_\calS(\bX)$, $\calI$, $\calJ$, $\calS$

\end{algorithmic}
\end{algorithm}

The subset random sampling scheme is straightforward but violates the common assumption in MC problems that each row and column must produce at least one sample. To the best of our knowledge, theoretical guarantees for reconstructing the original FTVGS when the samples are structurally missing in entire rows and columns have not been explored. Therefore, we then study subset random sampling from a theoretical perspective and provide a sufficient condition that ensures reconstruction of the original signal with high probability from the samples generated by subset random sampling.

The following lemma ensures that the rank of $\bX(\calI, \calJ)$ remains $r$ with high probability, which is crucial for the subsequent theoretical derivation. \cref{lem:IJ} also provides the lower bounds on the number of selected rows and columns.

\begin{lemma}
\label{lem:IJ}
    For an FTVGS $\bX \in \Complex^{N \times T}$ satisfying \cref{as_r} and \cref{as_mu}, let $\bX_R$ and $\bX_{RC}$ be obtained by Procedure \ref{proc:samp}. Then if
    $$ |\calI| \ge 3r\mu_1(\bX) \frac{\ln{(2r/\delta)}}{\epsilon^2}, |\calJ| \ge 3r\mu_2(\bX) \frac{\ln{(2r/\delta)}}{\epsilon^2} $$
    for some $0 < \delta < 1$ and $0 < \epsilon < 1$, $\mathrm{rank}(\bX_R) = r$ holds with probability at least $1-\delta$ and $\mathrm{rank}(\bX_{RC}) = r$ holds with probability at least $(1-\delta)^2$.
\end{lemma}

The $\bX_R = \bU_R \Sigma_R \bV^T_R$ is shown in \cref{eq:SVD_XR} and the SVD of $\bX_{RC}$ is 
\begin{equation}
\begin{aligned}
    \label{eq:XRC_SVD}
    \bX_{RC} &= \bU_R \Sigma_R (\bV_R(\calJ, :))^T = \bU_R \tilde{\bU}_{RC} \Sigma_{RC} \bV_{RC}^T \\ &= \bU_{RC} \Sigma_{RC} \bV_{RC}^T,
\end{aligned}
\end{equation}
where $\Sigma_R (\bV_R(\calJ, :))^T$ has the compact SVD $\tilde{\bU}_{RC} \Sigma_{RC} \bV_{RC}^T$ with $\tilde{\bU}_{RC} \in \Real^{r \times r}$ being an orthonormal matrix, $\bV_{RC} \in \Real^{r \times |\calJ|}$, and $\bU_{RC} = \bU_R \tilde{\bU}_{RC} $ is a matrix with orthogonal columns.

Next, \cref{lem:mu} shows how the properties of the original signal $\bX$ are converted to those of its submatrix $\bX(\calI, \calJ)$. The proof of \cref{lem:mu} is presented in APPENDIX \ref{proof:lem_mu}.

\begin{lemma}
\label{lem:mu}
    Under the assumption of \cref{lem:IJ}, if $\calI$ and $\calJ$ satisfy the inequality of \cref{lem:IJ}, then the following conditions hold with probability at least $(1-p)^2, p = r \left[ \frac{e^{-\eta}}{(1-\eta)^{1-\eta}} \right]^{\log r}$:
    $$ || \bU_{RC} ||_{2,\infty} \le \kappa(\bX) \sqrt{\frac{\mu_1(\bX) r}{(1-\eta) |\calI|}}, $$
    $$ || \bV_{RC} ||_{2,\infty} \le \frac{\kappa(\bX)}{1-\eta} \sqrt{\frac{\mu_2(\bX) Nr}{|\calI| |\calJ|}}, $$
    where $\kappa(\bX)$ is the condition number of $\bX$ and $\eta \in [0, 1)$. 
\end{lemma}

By \cref{lem:IJ} and \cref{lem:mu}, we present our main theorem, and its proof is presented in APPENDIX \ref{proof:thm_S}.

\begin{theorem}
\label{thm_S}
    For an FTVGS $\bX \in \Complex^{N \times T}$ satisfying \cref{as_r} and \cref{as_mu}, suppose that $\bX_{RC}$ and $\calS$ is derived from Procedure \ref{proc:samp}. Then if 
    \begin{align*}
        |\calS| \ge \frac{32 \beta (\kappa(\bX))^4 r^2 N}{(1-\eta)^3} \mu_1(\bX) \mu_2(\bX) \frac{|\calI|+|\calJ|}{|\calI|} \log^2(2n),
    \end{align*}
    where $\beta > 1$, $n = \max{\{|\calI|,|\calJ|\}}$, the $\bX$ can be recovered from $\calS$ with probability at least 
    $$ (1-\delta)^2 (1-p)^2 -\frac{6 \log{n}}{(|\calI|+|\calJ|)^{2\beta-2}} -n^{2-2\beta^{1/2}}. $$
\end{theorem}

\cref{lem:IJ} and \cref{thm_S} provide the conditions for reconstructing the original FTVGS from subset random sampled samples. Among the probabilities in \cref{thm_S}, $(1-\delta)^2$ is the probability that guarantees $\mathrm{rank}(\bX_{RC}) = \mathrm{rank}(\bX) = r$ and $(1-p)^2$ is the probability that guarantees the incoherence of $\bX_{RC}$. We summarize the differences between random sampling of MC, CCS, and subset random sampling in \cref{tab:diff}.

\begin{table*}[htbp]
    \centering
    \caption{Differences between three schemes for random sampling of FTVGS.}
    \begin{tabular}{lccc}
        \toprule
         &  Random sampling of MC    & CCS    & Subset random sampling \\
        \midrule
        Assumption & \multicolumn{2}{c}{rank-$r$, $\{ \mu_1, \mu_2 \}$-incoherence}   & rank-$r$, smooth, $\{ \mu_1, \mu_2 \}$-incoherence \\
        \midrule
        \makecell[l]{Complexity of\\ required samples} & \makecell[c]{$O(r^2 n_2 \log(n_2) )$ \cite{MC2021}\\ ($n_2 = \max{\{N, T \}}$) } & $O(r^2 n_2 \log^2(n_2) )$ \cite{CCS} & $O(r^2 N \log^2(n))$  \\
        \bottomrule
    \end{tabular}
    \label{tab:diff}
\end{table*}

Our proposed subset random sampling scheme restricts the sample indices within $\bX(\calI, \calJ)$, meaning that no samples will be drawn from unselected rows and columns. However, this comes at the cost of some reconstruction error, as detailed in \cref{sec:experiment}. Since the seminal work on the MC problem \cite{MC1}, the lower bounds on the number of randomly sampled entries have been relatively loose. Similarly, the lower bound provided in \cref{thm_S}, based on the approach in \cite{MC1}, is also loose, particularly when $N$ and $T$ are small.

\section{EXPERIMENTS}
\label{sec:experiment}

We validate \cref{thm_S} on a real-world dataset and propose a reconstruction algorithm tailored to the characteristics of subset random sampling. The details of the algorithm will be analyzed in our subsequent work.

The traffic data METR-LA is collected from loop detectors in the highway of Los Angeles County \cite{data_traffic}. We chose data collected by 207 sensors for this experiment. Each sensor is regarded as a vertex in $\calV_\calG$ and we denote the data of every 512 consecutive timesteps as $\bX$ for simulation. The METR-LA is divided into 100 FTVGS in total. That is, each $\bX$ is a FTVGS with $N = 207$ vertices and $T = 512$ timesteps. 

During the sampling stage, we follow the \cref{proc:samp}. We reconstruct the original FTVGS by solving the optimization problem 
\begin{equation}
    \begin{aligned}
        &\min_{\bF_\calG,\! \bF_\calT, \! \bE} \! \sum_i g(\lambda_i (\hat{\bX}^T \hat{\bX})) \! + \! \gamma_\calG ||\bW_\calG \odot \bF_\calG||_1 \! + \! \gamma_\calT ||\bW_\calT \odot \bF_\calT||_1 \\
        & \begin{array}{l@{\quad}l@{\quad}l}
        \text{s.t.} & \hat{\bX} = \Psi_\calG \bF_\calG, \\
             & \hat{\bX}^T = \Psi_\calT \bF_\calT, \\
             & \scP_\calS(\bX) = \hat{\bX} + \bE, \scP_\calS(\bE) = 0,
        \end{array}
    \end{aligned}
\label{eq:optm}
\end{equation}
where $\odot$ denotes the element-wise multiplication of two matrices, $\lambda_i (\hat{\bX}^T \hat{\bX})$ is the $i$-th eigenvalue of $\hat{\bX}^T \hat{\bX}$, $g(x) = \log(x^{1/2}+1)$, and $||\cdot||_1$ represents the $\ell_1$-norm of the matrix. The two regularization terms cause each column of $\bF_\calG$ and $\bF_\calT$ to tend towards sparsity, which aligns with the assumption of bandlimited spectra in both the graph domain and the time domain. $\bW_\calG$ and $\bW_\calT$ signify the weighting matrices for weighted sparsity terms. $\gamma_\calG$ and $\gamma_\calT$ are regularization parameters. $\bE$ is an error matrix.

The normalized root mean square error (NRMSE) is used to describe the error between signals $\bX_a$ and $\bX_b$
$$ {\rm NRMSE}(\bX_a, \bX_b) = \frac{||\bX_a - \bX_b||_F}{||\bX_a||_F}. $$
For each $\bX$, we sample at different ratios, where $\rho_{RC}$ represents the sampling ratio of randomly selected rows or columns, \textit{i.e.}, 
$$|\calI| = \frac{{\rm Round}(\rho_{RC} N)}{N}, |\calJ| = \frac{{\rm Round}(\rho_{RC} T)}{T},$$ where ${\rm Round}(\cdot)$ represents rounding numbers to the nearest integer. $\rho_{\bX_{RC}}$ represents the ratio of uniform sampling on $\bX_{RC}$, \textit{i.e.}, 
$|\calS| = {\rm Round}(\rho_{\bX_{RC}} |\calI| |\calJ|) / ( |\calI| |\calJ|)$. And $\rho_{total} = |\calS|/(NT)$ is the total sampling ratio.
Under different sampling ratio settings, the NRMSE for the three reconstruction methods is calculated. The mean NRMSE of the 100 FTVGS are presented in \cref{tab:nrmse}.

\begin{table}[htbp]
    \centering
    \caption{Comparison of NRMSE for reconstruction methods under different sampling ratio settings.}
    \begin{tabular}{l|ccc}
        \toprule
        $\rho_{RC}$ &  $90\%$    & $80\%$    & $60\%$ \\
        $\rho_{\bX_{RC}}$ & $90\%$    & $80\%$    & $60\%$ \\
        $\rho_{total}$ &    $72.81\%$ & $51.37\%$ & $21.55\%$  \\
        \midrule
        SVT \cite{SVT}     & 0.4542 & 0.6315 & 0.8421 \\
        TNNR \cite{TNNR}   & 0.4390 & 0.6006 & 0.8103 \\
        LIMC \cite{TV2018} & 0.1199 & 0.1751 & 0.3531 \\
        Ours  & \textbf{0.1171} & \textbf{0.1662} & \textbf{0.3442} \\
        \bottomrule
    \end{tabular}
    \label{tab:nrmse}
\end{table}

From the results in \cref{tab:nrmse}, we can see that traditional MC methods are not very effective. Because these methods are based on the low-rank property and can accurately fill in the missing positions when the missing elements in the FTVGS follow a random distribution. However, the samples obtained through our subset random sampling violate their assumptions. The LIMC and our method (\cref{eq:optm}) take into account the correlations between the rows and columns, demonstrating effectiveness and superiority in reconstructing the FTVGS from samples obtained through subset random sampling. This validates the conclusions of \cref{thm_S}.


\section{Conclusions}

In this work, based on the needs of practical applications, we propose a subset random sampling scheme for FTVGS, which involves uniform sampling from the submatrix $\bX_{RC}$ instead of the entire $\bX$. Theoretically, we proved sufficient conditions that guarantee the reconstruction of the original FTVGS with high probability under the subset random sampling scheme. Practically, we validated the effectiveness of reconstructing the original FTVGS from samples on a real-world dataset. Furthermore, for samples with randomly missing elements and some rows and columns are also randomly missing, we will consider targeted reconstruction algorithms to further reduce the NRMSE.

\section*{Acknowledgment}
This work was supported in part by the Fundamental R$\&$D Program of State Key Laboratory of Integrated Chips and Systems Under Grant KVH2327002.

\appendices

\section{Proof of Lemma \ref{lem:IJ}}
\label{proof:lem_IJ}

From the SVD of $\bX$ we know that $\bU \in \Real^{N \times r}$ is a matrix with orthogonal columns. Selecting the rows of $\bX$ is essentially row sampling on $\bU$.
    
    Theorem 4.1 and Corollary 4.2 in \cite{rowsample} proved that $\mathrm{rank}(\bU(\calI, :)) = r$ holds with probability at least $1-\delta$ if
    $$ |\calI| \ge 3N \xi_1 \frac{\ln{(2r/\delta)}}{\epsilon^2} $$
    for some $0 < \delta < 1$ and $0 < \epsilon < 1$, where $\xi_1 = || \bU ||_{2, \infty}^2$.

    According to \cref{as_mu}, $ \mu_1(\bX) r/N \ge || \bU ||_{2, \infty}^2 = \xi_1$, so we have 
    $$ |\calI| \ge 3r\mu_1(\bX) \frac{\ln{(2r/\delta)}}{\epsilon^2}. $$
    Further, we have $\bX_R = \bU(\calI, :) \Sigma \bV^T$, where $\Sigma \bV^T$ is a matrix with orthogonal rows. Then $\mathrm{rank}(\bX_R) = r$ holds with probability at least $1-\delta$.

    
    The SVD of $\bX_R$ is 
    \begin{equation}
    \label{eq:SVD_XR}
        \bX_R = \bU(\calI, :) \Sigma \bV^T = \bU_R \Sigma_R \tilde{\bV}^T_R \bV^T = \bU_R \Sigma_R \bV^T_R,
    \end{equation}
    where $\bU(\calI, :) \Sigma$ has the compact SVD $\bU_R \Sigma_R \tilde{\bV}^T_R$ with $\tilde{\bV}_R \in \Real^{r \times r}$ being an orthonormal matrix, and $\bV_R = \bV \tilde{\bV}_R $.
    
    The same argument yields
    $ |\calJ| \ge 3T \xi_2 \ln{(2r/\delta)}/\epsilon^2$
    for some $0 < \delta < 1$ and $0 < \epsilon < 1$, where $\xi_2 = || \bV_R ||_{2, \infty}^2 $ \cite{rowsample}.
    
    We have
    \begin{equation}
    \label{eq:L2}
        \xi_2 = || \bV_R ||_{2, \infty}^2 = || \tilde{\bV}_R^T \bV ||_{2, \infty}^2  = || \bV ||_{2, \infty}^2 \le \frac{\mu_2(\bX) r}{T},
    \end{equation}
    so we get 
    $$ |\calJ| \ge 3r \mu_2(\bX) \frac{\ln{(2r/\delta)}}{\epsilon^2}. $$

    Since $\bU(\calI, :) \Sigma$ is a matrix with orthogonal columns with probability at least $1-\delta$. Then $\mathrm{rank}(\bX_{RC}) = r$ holds with probability at least $(1-\delta)^2$.

\section{Proof of Lemma \ref{lem:mu}}
\label{proof:lem_mu}

According to the SVD of $\bX_R$, Lemma 4.2 and Lemma 4.3 in \cite{rCUR} give
    $$ || \bU_R ||_{2, \infty} \! \le \! \kappa(\bX) \frac{|| \bX^\dagger_R ||_2}{|| \bX^\dagger ||_2} \sqrt{\frac{\mu_1(\bX)r}{N}},  \frac{|| \bX^\dagger_R ||_2}{|| \bX^\dagger ||_2} \! \le \! || \bU(\calI, :)^\dagger ||_2. $$ 
    Then we have
    \begin{equation*}
    \begin{aligned}
        || \bU_R ||_{2, \infty} &\le \kappa(\bX) || \bU(\calI, :)^\dagger ||_2 \sqrt{\frac{\mu_1(\bX)r}{N}} \\
        & = \kappa(\bX) \frac{1}{\sigma_{\mathrm{min}}(\bU(\calI, :))} \sqrt{\frac{\mu_1(\bX)r}{N}}. 
    \end{aligned}
    \end{equation*}
    According to the Lemma 3.4 in \cite{subsamp_sigma}, $ \sqrt{\frac{(1-\eta) |\calI|}{N}} \le \sigma_{\mathrm{min}}(\bU(\calI, :)) $
    with failure probability at most $p = r\left[ \frac{e^{-\eta}}{(1-\eta)^{1-\eta}} \right]^{\log r},\eta \in [0,1)$.
    Then
    \begin{equation*}
    \begin{aligned}
        || \bU_R ||_{2, \infty} &\le \kappa(\bX) \sqrt{\frac{N}{(1-\eta) |\calI|}} \sqrt{\frac{\mu_1(\bX)r}{N}} \\
        & = \kappa(\bX) \sqrt{\frac{1}{1-\eta}} \sqrt{\frac{\mu_1(\bX)r}{|\calI|}}
    \end{aligned}
    \end{equation*}
    holds with probability at least $1- p$.
    The \cref{eq:XRC_SVD} yields
    \begin{align*}
        || \bU_{RC} ||_{2,\infty} &= || \bU_R \tilde{\bU}_{RC} ||_{2, \infty} \\
        &= || \bU_R ||_{2, \infty} \le \kappa(\bX) \sqrt{\frac{\mu_1(\bX) r}{(1-\eta) |\calI|}}.
    \end{align*}
    
    Notice that $\bV_{RC}^T = \Sigma_{RC}^{-1} \bU_{RC}^T \bX_{RC}$. Thus for any $i$,
    \begin{equation}
    \label{eq:VRC}
        ||\bV_{RC}^T \be_i||_2 \le \frac{1}{\sigma_{\mathrm{min}}(\bX_{RC})} ||\bX_{RC} \be_i||_2,
    \end{equation}
    where $\be_i$ is a unit vector whose $i$-th element is 1.
    Since $\bX_{RC} = \bX(:, \calJ) = \bU_R \Sigma_R (\bV_R(\calJ,:))^T$,
    \begin{equation}
        \label{eq:XRC}
        \begin{aligned}
            ||\bX_{RC} \be_i||_2 &\le \sigma_1(\bU_R) \sigma_1(\bX_R) ||(\bV_R(\calJ,:))^T \be_i||_2 \\
        &= \sigma_1(\bX_R) ||(\bV_R(\calJ,:))^T \be_i||_2.
        \end{aligned}
    \end{equation}
    Then combining (\ref{eq:VRC}) and (\ref{eq:XRC}), we have
    \begin{align*}
        || \bV_{RC} ||_{2,\infty} &\le \frac{\sigma_1(\bX_R)}{\sigma_{\mathrm{min}}(\bX_{RC})} ||(\bV_R(\calJ,:)) ||_{2,\infty} \\
        & \overset{(a)}{\le} \frac{\sigma_1(\bX_R)}{\sigma_{\mathrm{min}}(\bX_{RC})} || \bV_R ||_{2,\infty} \\
        & = \kappa(\bX) \frac{\sigma_{\mathrm{min}}(\bX)}{\sigma_{\mathrm{min}}(\bX_{RC})} || \bV_R ||_{2,\infty} \\
        & = \kappa(\bX) \frac{||\bX_{RC}^\dagger||_2}{||\bX^\dagger||_2} || \bV_R ||_{2,\infty} \\
        & \overset{(b)}{\le} \kappa(\bX) \frac{1}{\sigma_{\mathrm{min}}(\bV(\calJ,:)) \sigma_{\mathrm{min}}(\bU(\calI,:))} || \bV_R ||_{2,\infty} \\
        & \overset{(c)}{\le} \frac{\kappa(\bX)}{1-\eta}  \sqrt{\frac{NT}{|\calI| |\calJ|}} || \bV_R ||_{2,\infty} \\
        &\le \frac{\kappa(\bX)}{1-\eta}  \sqrt{\frac{NT}{|\calI| |\calJ|}} \sqrt{\frac{\mu_2(\bX) r}{T}}
    \end{align*}
    with probability at least $(1- p)^2$, where Lemma 4.2 in \cite{rCUR} gives (a) and Lemma 3.4 in \cite{subsamp_sigma} gives (c).
    For (b), we have $ \bX_{RC} = \bU(\calI,:) \Sigma (\bV(\calJ,:))^T$, so 
    $$ ||\bX_{RC}^\dagger||_2 \le ||((\bV(\calJ,:))^T)^\dagger||_2 ||\bX^\dagger||_2 ||(\bU(\calI,:))^\dagger||_2. $$

\section{Proof of Theorem \ref{thm_S}}
\label{proof:thm_S}

    Since $\calI$ and $\calJ$ are chosen uniformly from $\calV_\calG$ and $\calV_\calT$ based on \cref{proc:samp}, according to \cref{lem:mu} we have
    \begin{align*}
        ||\bU_{RC} \bV_{RC}^T||_\infty &\le || \bU_{RC} ||_{2,\infty} || \bV_{RC} ||_{2,\infty} \\
        &\le \frac{(\kappa(\bX))^2}{(1-\eta)^{3/2}} \sqrt{\frac{\mu_1(\bX) \mu_2(\bX) Nr^2}{|\calI|} } \sqrt{\frac{r}{|\calI| |\calJ|}}
    \end{align*}
    for $\eta \in [0, 1)$ with probability at least $(1-p)^2$.

    By Theorem 2 in \cite{MC2}, we obtain that if 
    \begin{equation}
        \label{eq:S}
        |\calS| \ge \frac{32 \beta (\kappa(\bX))^4 r^2 N}{(1-\eta)^3} \mu_1(\bX) \mu_2(\bX) \frac{|\calI|+|\calJ|}{|\calI|} \log^2(2n)
    \end{equation}
    for some $\beta > 1$, $\bX_{RC}$ can be reconstructed with probability
    $$ 1 - \frac{6 \log n}{(|\calI|+|\calJ|)^{2\beta-2}} - n^{2-2\beta^{1/2}}. $$

    Based on \cref{as_r}, we know that $\parallel \triangledown \bX(i,j) \parallel < C$, so $|\bQ^T \bX(i,j)| < \infty$ and $\bX \bD_1(i,j) < \infty$. We define that 
    $$\parallel \triangledown \bX(i,j) \parallel_a = \sqrt{(\bQ^T \bX(i,j))^2 + (\bX \bD_1(i,j))^2 + a^2} $$
    for $0 < a < \infty$, and $0 < \parallel \triangledown \bX(i,j) \parallel_a < \infty$.
    Similarly, we have
    $$ \left| \bQ^T \bX(i,j) \right|_a = \sqrt{(\bQ^T \bX(i,j))^2 + a^2} < \infty, $$
    $$ |\bX \bD_1(i,j)|_a = \sqrt{(\bX \bD_1(i,j))^2 + a^2} < \infty. $$
    According to Section 5 in \cite{XRC2X}, we have
    \begin{align*}
        & \triangledown \! \left( \frac{\triangledown \bX(i,j)}{\parallel \! \triangledown \bX(i,j) \! \parallel_a} \right) \! = \! \frac{1}{\parallel \! \triangledown \bX(i,j) \! \parallel_a^3} ( |\bX \bD_1(i,j)|_a^2 \bX(:,j)^T \bL \bX(:,j) \\
        & +\! \left| \bQ^T \bX(i,j) \right|_a^2 \bX \bD_2(i,j) \! + \! 2 \bQ^T \bX(i,j) \bX \bD_1(i,j) \bQ^T \bX \bD_1(i,j) ) 
    \end{align*}
    and $\left| \triangledown \left( \frac{\triangledown \bX(i,j)}{\parallel \triangledown \bX(i,j) \parallel_a} \right) \right| < \infty$.
    Thus, we can reconstruct $\bX$ from $\bX_{RC}$ by solving the TV inpainting model \cite{XRC2X}. Then if \cref{eq:S} holds, we are able to reconstruct $\bX$ from $\calS$ with probability at least 
    $$ (1-\delta)^2 (1-p)^2 -\frac{6 \log{n}}{(|\calI|+|\calJ|)^{2\beta-2}} -n^{2-2\beta^{1/2}}. $$

\bibliographystyle{IEEEtran}
\bibliography{references}

\end{document}